\documentclass[%
 reprint,
showpacs,
 amsmath,amssymb,
 aps,
]{revtex4-1}

\usepackage{graphicx}
\usepackage{dcolumn}
\usepackage{bm}
\usepackage{textcomp}

\usepackage[textsize=small]{todonotes}       
\usepackage{textcomp}
\usepackage{mathrsfs}

\usepackage{color}

\newcommand {\be} {\begin{equation}}

\newcommand {\bea}  {\begin{eqnarray}}
\newcommand {\eea}  {\end{eqnarray}}

\newcommand {\ee} {\end{equation}}
\newcommand {\bdm} {\begin{displaymath}}
\newcommand {\edm} {\end{displaymath}}
\newcommand {\ba}  {\begin{array}}
\newcommand {\ea}  {\end{array}}

\newcommand{\cla}[1]{\textcolor[rgb]{0,0,0}{#1}}

\newcommand{\ah}{\`a}

\begin{document}

\preprint{APS/123-QED}

\title{$O(N)$ fluctuations and lattice distortions in 1-dimensional systems}


\author{Claudio Giberti}
\email{claudio.giberti@unimore.it}
\affiliation{Dipartimento di Scienze e Metodi dell'Ingegneria,
Universit\`a di Modena e Reggio E., \\ Via Amendola 2, Padiglione Morselli, I-42122 Reggio E., Italy.}
\author{Lamberto Rondoni}
\email{lamberto.rondoni@polito.it}
\affiliation{Dipartimento di Matematica, Politecnico di Torino,\\
Corso Duca degli Abruzzi 24 I-10129 Torino, Italy.}
\author{Cecilia Vernia}
\email{cecilia.vernia@unimore.it}
\affiliation{Dipartimento di Scienze Fisiche Informatiche e Matematiche, Universit\`a  di Modena 
e Reggio E.,\\
Via Campi 213/B, I-41125 Modena, Italy\\}

\date{\today}

\begin{abstract}
Statistical mechanics harmonizes mechanical and thermodynamical quantities, via the 
notion of local thermodynamic equilibrium (LTE). In absence of external drivings, LTE 
becomes equilibrium tout court, and states are characterized by several thermodynamic 
quantities, each of which is associated with negligibly fluctuating
microscopic properties. Under small driving and LTE, locally conserved quantities are 
transported as prescribed by linear hydrodynamic laws, in which the local material 
properties of the system are represented by the transport coefficients.
In 1-dimensional systems, on the other
hand, the transport coefficients often appear to depend on the 
global state, rather than on the local state of the system at hand. We interpret these facts within the framework
of boundary driven 1-dimensional Lennard-Jones chains of $N$ oscillators, observing that they
experience non-negligible $O(N)$ lattice distortions and fluctuations. This implies that standard 
hydrodynamics and certain expressions of energy flow do not apply in these cases. One possible 
modification of the energy flow is considered.
\end{abstract}


\maketitle

\section{Introduction}
In a seminal paper, Rieder, Lebowitz and Lieb investigated the properties of chains of $N$ 
harmonic oscillators, interacting at their ends with stochastic heat baths \cite{Lebow}.
These authors proved that while energy flows from hot to cold baths, the kinetic 
temperature profile decreases exponentially in the direction of the hotter bath, rather 
than increasing, and in the bulk its slope vanishes as $N$ grows. Thus, in case the kinetic 
temperature equals the thermodynamic temperature, heat flows against
the direction of energy, in the bulk of such 1D systems. Moreover, 
no steady state is reached, because at boundaries, heat flows in the opposite direction.
Taken as a paradox without explanation in Ref.\cite{Lebow}, this fact reveals that, in 
harmonic chains of oscillators, the kinetic temperature does not correspond to the 
thermodynamic temperature, or the energy flux does not represent a heat flux, or both.
Unexpected phenomena that seem to contradict the hydrodynanmic laws of transport, {\em e.g.}\  
currents going against the density gradient,  
a phenomenon called ``uphill diffusion'',  can be observed in several experimental settings;
see  e.g.\ Ref.\cite{ColGiaGibVer} for further references and a nonequilibrium model 
with phase transition exhibiting uphill diffusion. However, the thermodynamic relevance of 
such models is still under investigation.

As a matter of fact, temperature and heat pertain to macroscopic objects with microscopic states corresponding
to Local Thermodynamic Equilibrium (LTE); 
they cannot be directly identified with mechanical quantities such as kinetic energy and
energy flux, Ref.\cite{LandauV} \S 9, \cite{Chibbaro} Chapters 3, 4 
and 5. LTE is the essence of Thermodynamics: it can be viewed at once as the precondition 
for the existence of the thermodynamic fields, such as temperature and heat, and as the 
natural state of objects obeying the thermodynamic laws. The microscopic conditions under 
which LTE is expected to hold are extensively discussed in the literature, {\em e.g.}\ 
\cite{Prigogine} Section 15.1, \cite{Spohn} Section 
2.3, \cite{Bellissard} Section 3.3, \cite{Kreuzer} Chapter 1. In short, LTE requires
the existence of three well separated time and space scales, so that: {\bf 1.}\ a macroscopic 
object can be subdivided in mesoscopic cells that look like a point to macroscopic 
observers, while containing a large number of molecules; {\bf 2.}\ boundary effects
are negligible compared to bulk effects, so that the contributions of neighboring cells 
to the mass and energy of a given cell are inappreciable within a cell; 
{\bf 3.}\ particle interactions allow the cells to thermalize (positions and velocities 
become respectively uniformly and Maxwell-Boltzmann distributed) within times that 
are mere instants on the macroscopic scale. 

That macroscopic observables are not affected by microscopic fluctuations, despite the exceedingly 
disordered and energetic microscopic motions, is essential for mesoscopic quantities to be sufficiently 
stable that thermodynamic laws apply, {\em e.g.}\ \cite{LandauV}, \S 1 and \S 2. This is the case for 
a quantity that is spatially weakly inhomogeneous, when the number $N$ of particles in a cell 
is large, and the molecular interactions randomize positions and momenta so that, for 
instance, the fluctuations of a quantity $\phi$ of size $O(N)$ are order $O(\sqrt{N})$. 
The bulk of the cell then dominates in- and out-fluxes, and variations of $\phi$ 
are sufficiently slow on the mesoscopic scale. Quantitatively, the space and time scales 
for which this description holds depend on the properties of the microscopic components 
of the systems of interest, \cite{Prigogine,Kreuzer,Spohn,Bellissard,Pigo1,Pigo2}. 

Under the LTE condition, matter can be considered a continuum, obeying hydrodynamic laws,
{\em i.e.}\ balance equations for locally conserved quantities, such 
as mass, momentum and energy \cite{DeGroot,CR98,RC02}. For small to moderate driving, 
they take a linear form, in which the local material properties are expressed
by the linear transport coefficients. Locality implies that such 
coefficients do not depend on the conditions of the system far away from 
the considered region. The thermal conductivity of an iron bar at a given temperature at a given 
point in space does not depend on the conditions of the bar far from that region; cutting the bar 
in two, or joining it to another bar, without changing the local state, leaves unchanged its local 
properties.

Fluctuations remain of course present in systems made of particles; they are larger for larger systems, 
they may be observed \cite{AURIGA1,AURIGA2}, and they play a major role 
in many circumstances, see {\em e.g.}\ Refs.\cite{Mish,HickMish}. This
motivates a considerable fraction of research in statistical physics, {\em e.g.}\ \cite{StochTherm,CasatiRev},
concerning scales much smaller than the macroscopic ones, or occurring in 
low dimensional (1D and 2D) systems \cite{Li1,Dhar,Li2,LepriEd,CasatiJ}. In these phenomena, the linear transport 
coefficients do not always 
seem to exist \cite{LepriEd}, the robustness and universality of the thermodynamic laws appear 
to be violated, and the behaviours appear to be strongly affected by boundary conditions and by all parameters 
that characterize a given object \cite{ExpAnomFourier,JeppsR,GibRon,Zhao1,Zhao2,onorato,Saito}. 
It is also well known that chains of oscillators behave more like some kind of 
(non-standard) fluids than like solids, because of the loss of crystalline structure, caused by 
cumulative position fluctuations \cite{Peierls}. Consequently, a fluid-like (possibly fluctuating) 
description has been adopted in a number of papers, cf.\ Refs.\cite{NaRam02,MaNa06}.

In driven systems, the situation is problematic also because equipartition may be violated \cite{Breaking,Sengers,PSV17},
the statistic describing the state of the system is model dependent, and the ergodic properties of the particles dynamics are only partially understood \cite{GGnoneq,EWSR2016}. Hence, 
there is no universally accepted microscopic notion of nonequilibrium temperature \cite{MR,Jou1,Jou2,JouBook,Criado,He,PSV17}. Further, a microscopic definition of heat flux requires a clear distinction between energy transport due to macroscopic motions (convection), and transport without macroscopic motions (conduction), cf.\ Chapter 4 of Ref.\cite{Zema}, and Section III.2 and Chapter XI of Ref.\cite{DeGroot}. In 1D systems, this may not be always possible \cite{inpreparat}.

One possible interpretation of these facts is that LTE is violated in some situations, hence that 
thermodynamic concepts may be inappropriate \cite{JeppsR,GibRon}. Another interpretation is that 
thermodynamic notions should be modified to treat small and strongly nonequilibrium systems, see 
{\em e.g.}\ \cite{MR,Jou1,Jou2,He}.
It is therefore interesting to investigate the validity and universality of the mechanical counterparts 
of thermodynamic quantities, in situations in which LTE is not expected to hold, and ``anomalous'' 
phenomena have been reported. 

We address such questions considering chains of $N$ Lennard-Jones oscillators interacting with 
deterministic baths at their ends, and without on-site potentials. Our central findings are that:
\begin{itemize}
\item
thermostats at different temperatures induce $O(N)$ distortions of the 
equilibrium lattice, resulting in highly in-homogeneous chains;
\item
thermostats induce collective order $O(N)$ fluctuations, 
{\em i.e.}\ ``macroscopic'' motions. Negligible incoherent $O(\sqrt{N})$ vibrations typical of 
3D equilibrium systems are thus replaced by kind of convective motions, even in 
chains bounded by still walls.
\end{itemize}

These observations combined with the results of Ref.\cite{Lebow} and further literature, 
{\em e.g.}\ Refs.\cite{MR,Jou1,Jou2,JouBook,He,Breaking}, suggest 
that microscopic definitions appropriate for 3-dimensional equilibrium thermodynamic quantities, need 
extra scrutiny in 1D. In particular, we illustrate how $O(N)$ fluctuations and lattice distortions 
affect the collective behaviour of 1D systems, considering the notion of heat flux, $J$ 
say, given by Eq.(23) of Ref.\cite{LLP-PhyRep}. This way, we confirm from a different standpoint the conclusions
reached in previous studies on the inapplicability of standard hydrodynamics \cite{Hurtado,Politi}. We find 
that:
\begin{itemize}
\item
$J$ is not spatially uniform in steady states. Variations of $J$ decrease if the baths temperature difference is reduced 
at constant $N$, but they do not if the mean temperature gradient is reduced increasing $N$ at constant baths 
temperatures. 
\item
Dividing $J$ by the local 
mass density partially balances the lattice inhomogeneity and yields an approximately uniform quantity.
\end{itemize}
These observations should be combined with those of Refs.\cite{Politi,inpreparat}, according to which
collective and molecular motions are correlated, making hard to disentangle convection from conduction,
because single particles push their neighbors, producing kind of convective cascades.
That difficulties do not ease when $N$ grows, indicates 
that LTE, hence thermodynamic quantities, cannot be established in our 1D systems. We relate this fact 
with the $O(N)$ growth of fluctuations and lattice distortions.

\section{Chains of Lennard-Jones oscillators} \label{sectII}
Consider a 1D chain of $N$ identical moving particles of equal mass $m$, 
and positions $x_{i}$, $i=1,...,N$. Add two particles with fixed positions, $x_{0}=0$ and $x_{N+1}=(N+1)a$, where 
$a$ is the lattice spacing. Let nearest neighbors interact via the Lennard-Jones potential (LJ):
\be
V_{1}(r)=\epsilon \left [ \left (\frac{a}{r}\right)^{12} - 2  \left (\frac{a}{r}\right )^{6}  \right ],
\label{LJ}
\ee  
where $r$ is the distance between nearest neighbors: \cla{$r=|x_{i}-x_{i-1}|$} and $\epsilon>0$ is 
the depth of the potential well. Thus, $x_{i}=ai$, with 
$\, i=0,\ldots, N+1$, is a configuration of stable mechanical equilibrium for the system. We also 
consider interactions involving first and second nearest neighbors, with second 
potential given by  \cite{1st2ndpaper}:
\be
 V_{2}(s)=\epsilon \left [ \left (\frac{2a}{s}\right)^{12} - 2  \left (\frac{2a}{s}\right )^{6}  \right ],
\label{1st2nd}
\ee 
where \cla{$s=|x_{i}-x_{i-2}|$}. Further, we add two particles with fixed positions $x_{-1}=-a$ and $x_{N+2}=(N+2)a$.
With potential $V = V_{1}+V_{2}$, the system has the usual stable mechanical equilibrium configuration $x_{i}=ai$, 
$\, i=-1,\ldots, N+2$.
The first and last moving particles are in contact with two Nos\'e-Hoover thermostats, at kinetic temperatures 
$T_{L}$ (on the left) and $T_{R}$ (on the right) and with relaxation times $\theta_L$ and $\theta_R$. 
Introducing the forces
\be
\label{F1F2}
F_{1}(r)= \frac{\partial V_{1}}{\partial r }(r) ,\quad F_{2}(s)= \frac{\partial V_{2}}{\partial s }(s) \, ,
\ee 
the equations of motion are given by:
\bea
&&m \ddot{x}_{1} =  F_{1}(x_{1}) -   F_{1}(x_{2}-x_{1})-\xi_{1} \dot{x}_{1}, \\
&&m \ddot{x}_{i} =  F_{1}(x_{i}-x_{i-1}) - F_{1}(x_{i+1}-x_{i}), \, i=2,...,N-1, \;\;\;\;\; \\
&&m \ddot{x}_{N} =  F_{1}(x_{N}-x_{N-1}) -   F_{1}(x_{N+1}-x_{N})-\xi_{N} \dot{x}_{N},
\label{eqsmot}
\eea
with
\be
\dot{\xi}_{1} = \frac{1}{\theta_{L}^{2}}\left ( \frac{m \dot{x}_{1}^{2} }{T_{L}}-1\right ),\quad  \dot{\xi}_{N} = \frac{1}{\theta_{R}^{2}}\left ( \frac{m \dot{x}_{N}^{2} }{T_{R}}-1\right ),
\ee
in the case of nearest neighbors interaction. For first and second neighbors interactions, we have:
\begin{widetext}
\bea
&& \hskip -18pt  m \ddot{x}_{1} =  F_{1}(x_{1}) -   F_{1}(x_{2}-x_{1})  + F_{2}(x_{1}+a) -   F_{2}(x_{3}-x_{1}) -\xi_{1}\dot{x}_{1}, \nonumber \\
&& \hskip -18pt m \ddot{x}_{2} =  F_{1}(x_{2}-x_{1}) -   F_{1}(x_{3}-x_{2})  + F_{2}(x_{2}) -   F_{2}(x_{4}-x_{2})-\xi_{2}\dot{x}_{2},\nonumber  \\
&& \hskip -18pt m \ddot{x}_{i} =  F_{1}(x_{i}-x_{i-1}) -   F_{1}(x_{i+1}-x_{i})  + F_{2}(x_{i}-x_{i-2}) -   F_{2}(x_{i+2}-x_{i}),  \qquad i=3,\ldots, N-2, \\
&& \hskip -18pt m \ddot{x}_{N-1} = F_{1}(x_{N-1}-x_{N-2}) - F_{1}(x_{N}-x_{N-1})  + F_{2}(x_{N-1}-x_{N-3}) -   F_{2}(x_{N+1}-x_{N-1})-\xi_{N-1}\dot{x}_{N-1},\nonumber \\
&& \hskip -18pt m \ddot{x}_{N} = F_{1}(x_{N}-x_{N-1}) - F_{1}(x_{N+1}-x_{N})  + F_{2}(x_{N}-x_{N-2}) - F_{2}(x_{N+2}-x_{N})-\xi_{N}\dot{x}_{N},\nonumber 
\label{eqsmot2}
\eea
\end{widetext}
with
\be
\begin{split}
\dot{\xi}_{l}  &= \frac{1}{\theta_{L}^{2}}\left ( \frac{m \dot{x}_{l}^{2} }{T_{L}}-1\right ),\, 
l=1,2,\\
 \dot{\xi}_{l} &= \frac{1}{\theta_{R}^{2}}\left ( \frac{m \dot{x}_{l}^{2} }{T_{R}}-1\right ),\,  l=N-1,N.
\end{split}
\ee
The hard-core nature of the LJ potentials preserves the order of particles:
$0< x_{1}<x_{2}<\cdots <x_{N}<(N+1)a$ holds 
at all times, if it does at the initial time \cite{InSome}.

For such systems, a form of single particle virial relation is often found to hold \cite{Virial}. 
That fact is usually mentioned to identify the average kinetic energy of a given particle with the 
temperature $T_i$ in position $x_i$ \cite{LLP-PhyRep}:
\be
T_i=\left \langle {{p_i}^2 \over m}\right\rangle ~, \qquad i=1,...,N.
\label{kinT}
\ee
Here, $p_i$ is the momentum of particle $i$, the angular brackets ${\langle \cdot \rangle}$ denote 
time average, and $T_i$ is called {\em single particle kinetic temperature}. In the case in 
which $T_L \ne T_R$, the single particle kinetic temperature profile may take rather peculiar forms, 
compared to the linear thermodynamic temperature profiles in homogeneous solids when Fourier law holds.
This is illustrated in great detail in the specialized literature, cf.\ 
\cite{LepriEd,GibRon,MejiaPoliti,LLP-PhyRep,LLP,DharReview,MejiaPoliti} just to cite a few. 
Also, numerically simulated profiles of various kinds of 1D systems, appear to be
sensitive to parameters such as the relaxation constants of the thermostats, the interaction
parameters, the form of the boundaries {\em etc.}\ cf.\ {\em e.g.}\ Ref.\cite{GibRon}.
This is not surprising, since many correlations persist in space and time in low dimensional systems, 
hindering the realization 
of LTE \cite{JeppsR, Hurtado, Politi, RJepps, BiancaJR, Salari,inpreparat}, and leading to anomalous behaviors.

In the following sections, we report our results about systems with various numbers of particles $N$.  
The parameters defining the Lennard-Jones potentials are $\epsilon =1$ and $a=1$, while the mass of 
the particles is $m=1$. The relaxation times of the thermostats $\theta_L$ and $\theta_R$ are set to 
1. The numerical integrator used is the fourth-order Runge-Kutta method with step size $10^{-3}$. 
The time averages are typically taken over $O(10^8) - O(10^9)$ time steps in the stationary state. 

\section{Large lattice deformations and fluctuations} \label{fluid}
The distinction between the different states of aggregation of matter is not strictly possible in 1D systems 
with short range interactions; one nevertheless realizes that our oscillators chains are more similar to 
(a kind of) compressible fluids than to solids \cite{Hurtado,GibRon}. In particular, Ref.\cite{Politi} 
shows persistent correlations, $O(N)$ dependence of relaxation times, and the failure of standard hydrodynamics, 
in non-driven LJ systems.

\begin{figure}
\includegraphics[width=0.5\textwidth]{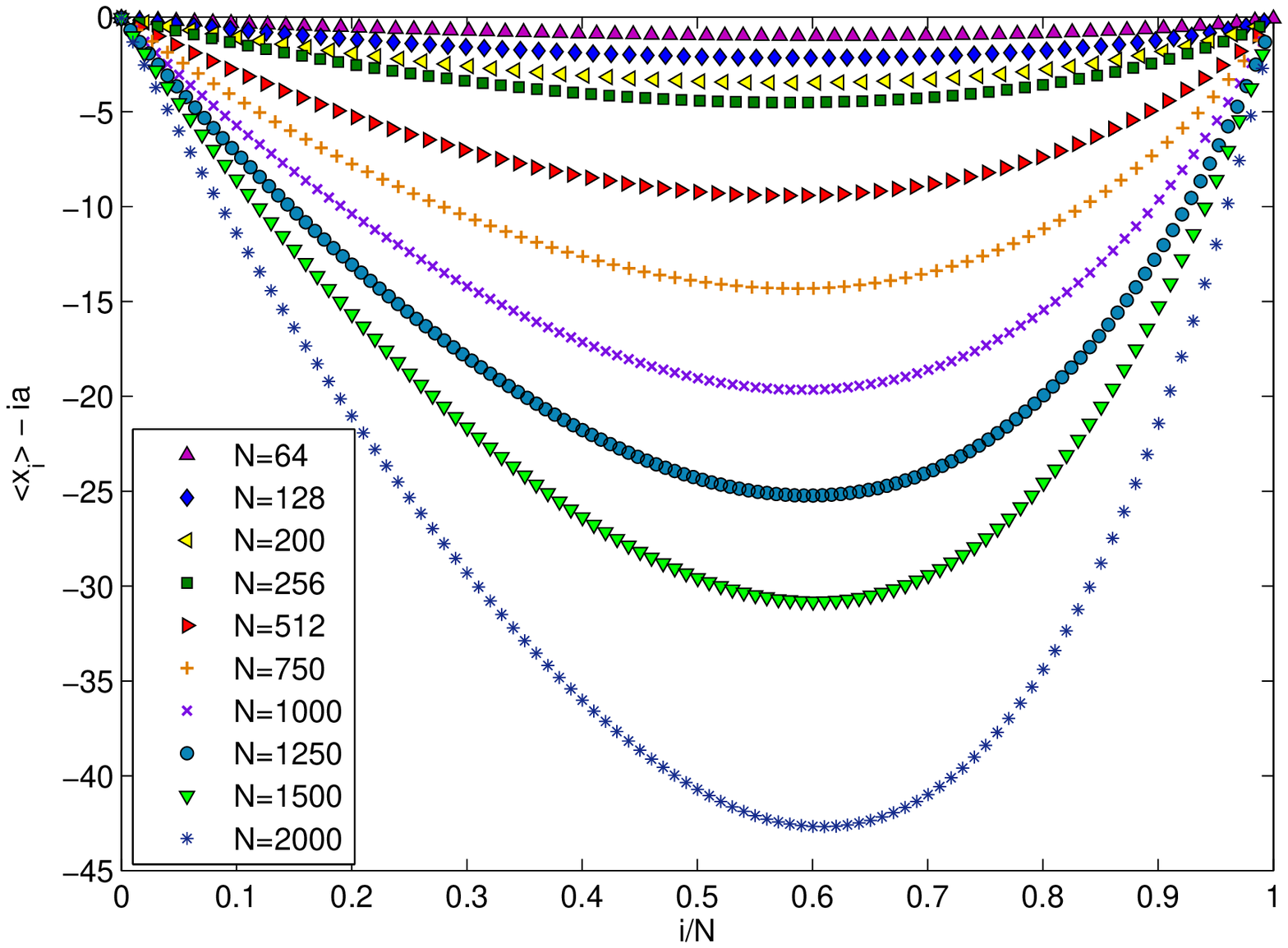} 
\includegraphics[width=0.5\textwidth]{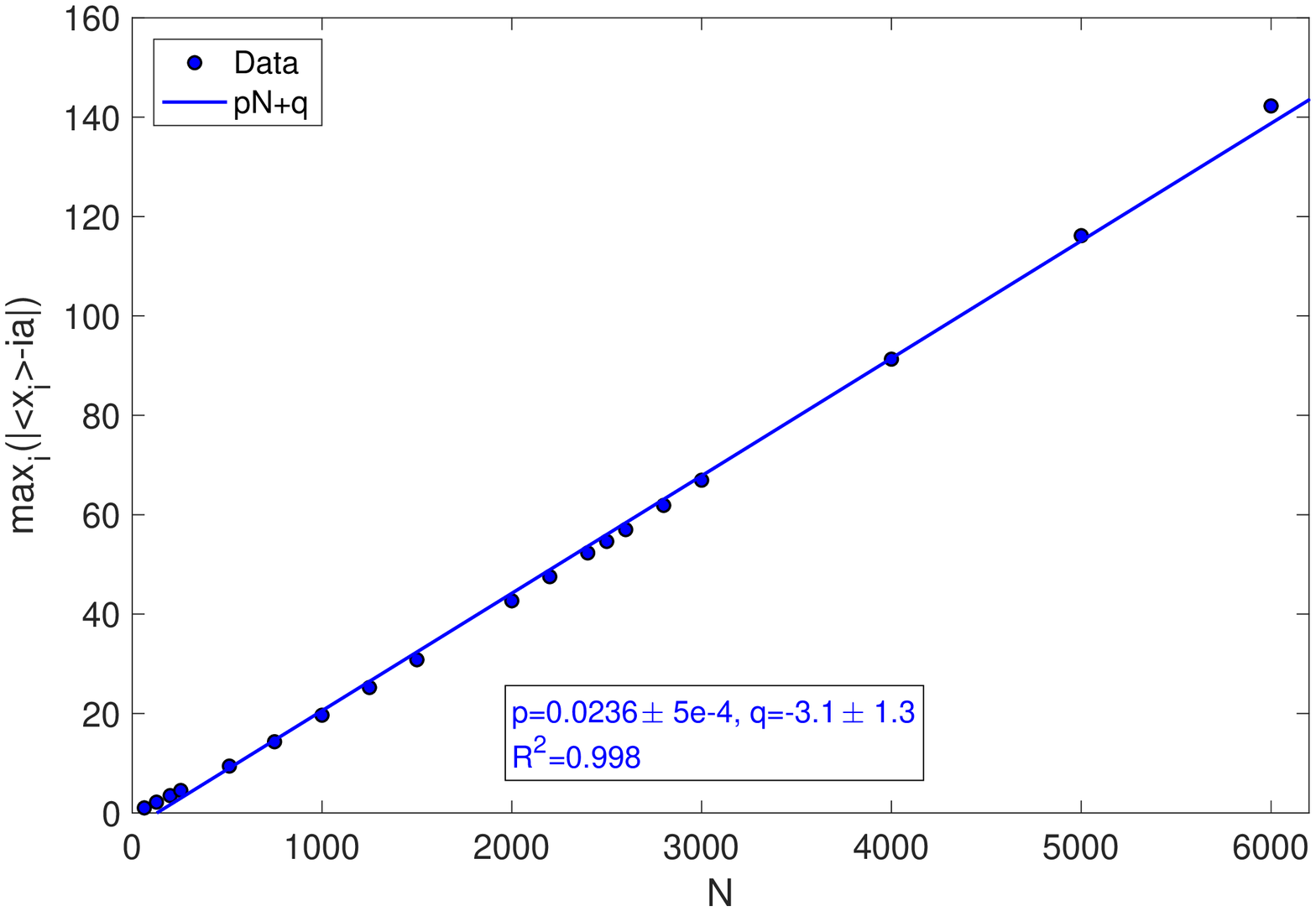}
\caption{\label{pag2June} Upper panel: plot of the displacement
of the mean position of particle $i$ from its mechanical equilibrium position, 
$(\langle x_i \rangle - i a)$, 
for different values of 
$N$, for first and second neighbors interaction, when $T_L=1$ and $T_R=10$. The lattice is strongly
distorted in presence of temperature differences. Lower panel: linear fit of  
$\max_i (\langle x_i \rangle -i a) $ as a function of $N$ ranging from 64 to 6000, for $N>400$. The label 
of the particle corresponding to the maximum lattice distortion is fitted by 
$i_{\mbox{\small mld}} = 0.6063 N - 6.804$ with $R^2=0.9997$.
}
\end{figure}

Along similar lines of investigation, we find that temperature differences 
at the boundaries of the chains induce ``macroscopic'' deformations of the periodic structure of the lattice. 
For all $i$, we obtain $(\langle x_i \rangle -i a) \sim O(N)$, as shown in Fig.\ref{pag2June}, whose lower 
panel plots the quantity $\max_i (\langle x_i \rangle -i a)$ as a function of $N$.

Our second observation is that the presence of thermostats at different temperatures enhances the size of the  
vibrations of each particle $i$ about its average position $\langle x_i \rangle$. Such vibrations are  
order $O(i^{1/2})$ in chains without thermostats with origin in $i=0$ \cite{Peierls}, which means that,
for sufficiently large $i$, position fluctuations are incompatible with a crystal structure. 
In our framework, the length of chains is bounded, therefore the size of particle vibrations cannot 
indefinitely grow with particle index $i$: the vibrations are larger for particles in the bulk than 
for particles near the boundaries of the chains. 

\begin{figure}
\includegraphics[width=0.5\textwidth]{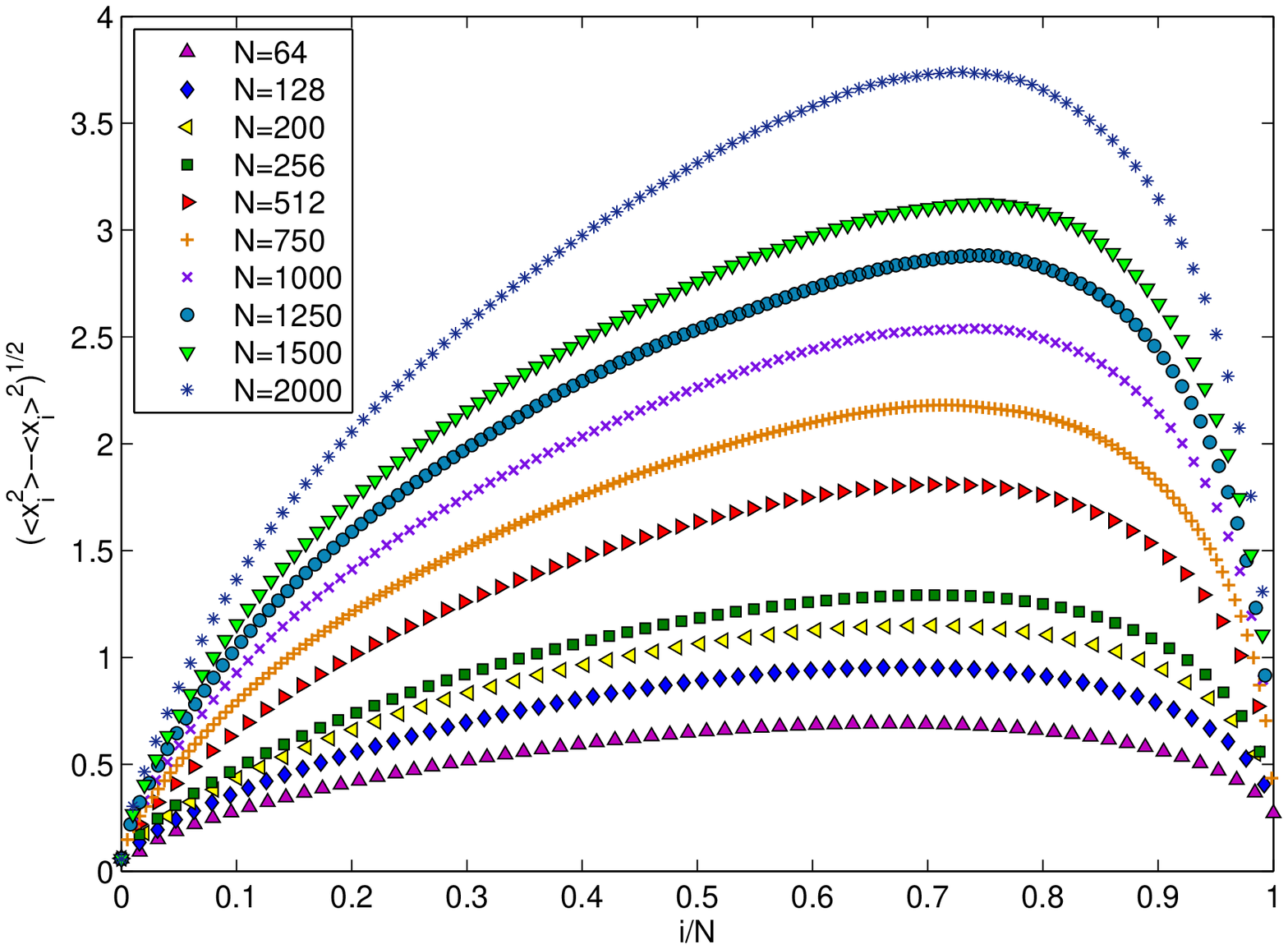} 
\includegraphics[width=0.5\textwidth]{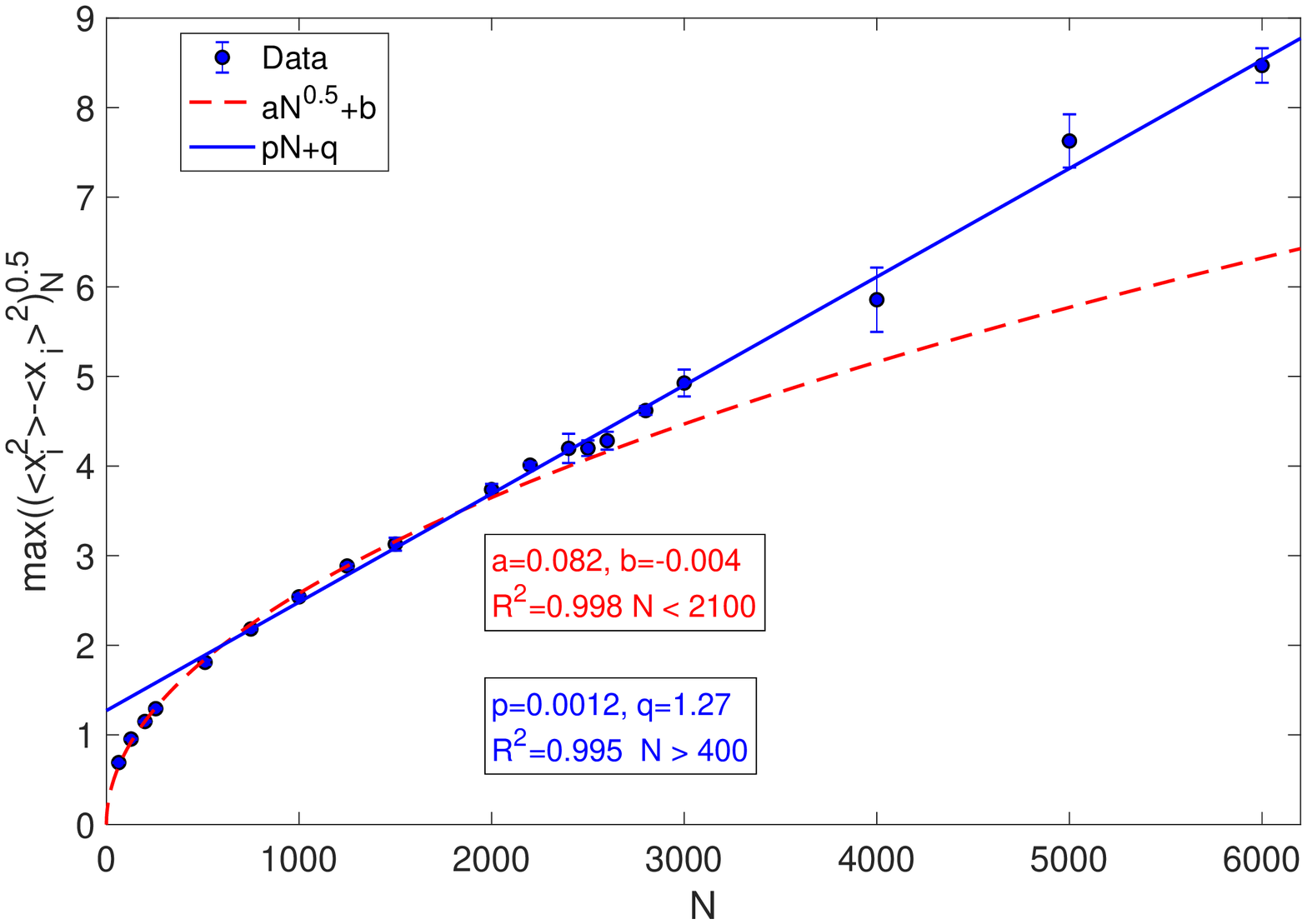}
\caption{\label{page67June} Upper panel: standard deviations of the particles vibrations 
about their average position, in lattice vectors units, for the case of Fig.\ref{pag2June}.
Lower panel: dependence on $N$ (ranging from 64 to 6000) of the maximum standard deviation together with a linear
fit for $N>400$ (continuous blue line)  and one square root fit for lattices with $N<2100$ (dashed red line). 
Growing linearly with $N$, collective vibrations look like convective motions. 
The label of the particle corresponding to the maximum fluctuation amplitude is fitted by 
$i_{\mbox{\small mfa}} = 0.7398 N - 6.75$ with $R^2=0.9993$.
}
\end{figure}

\begin{figure}
\includegraphics[width=0.5\textwidth]{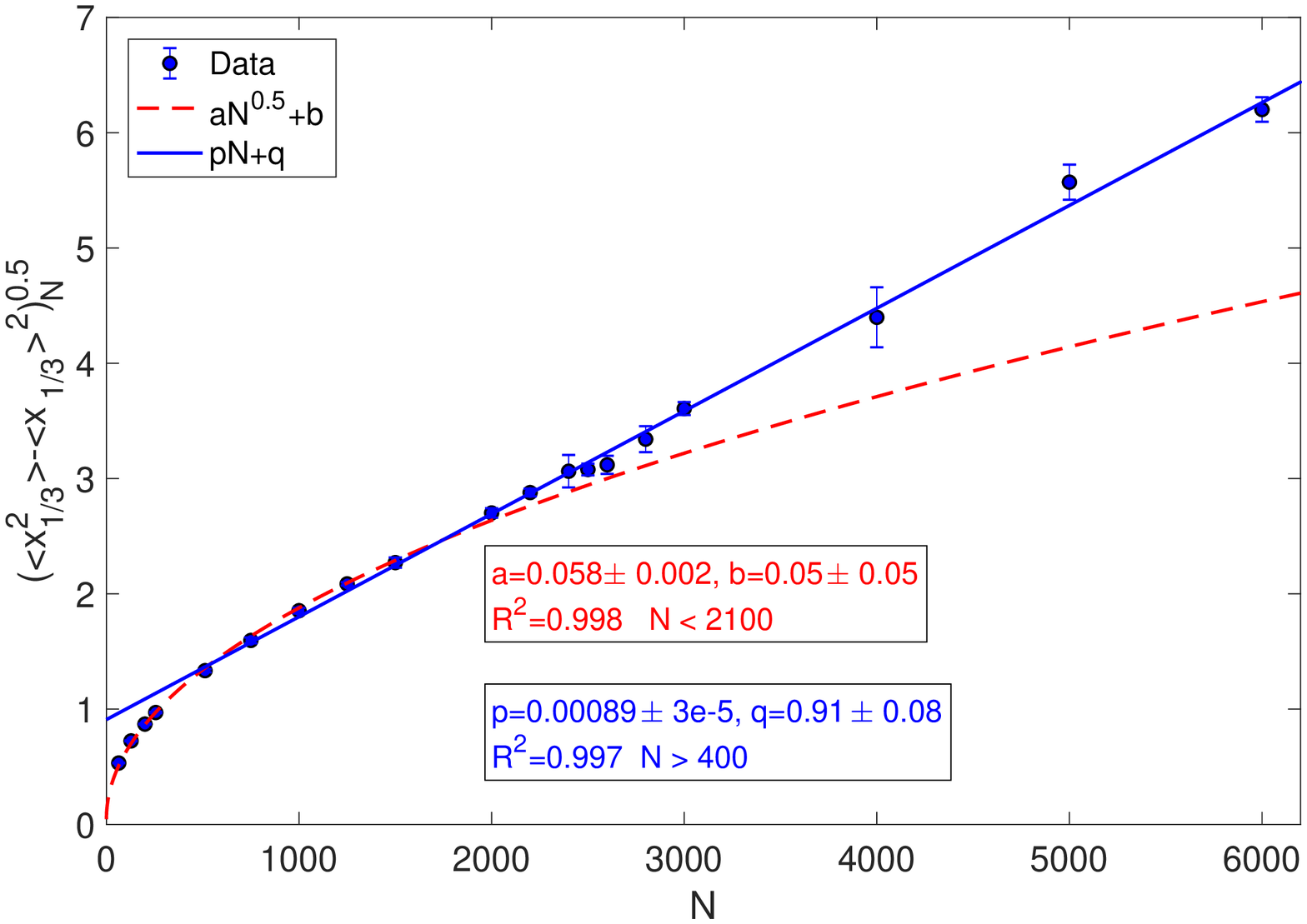} 
\includegraphics[width=0.5\textwidth]{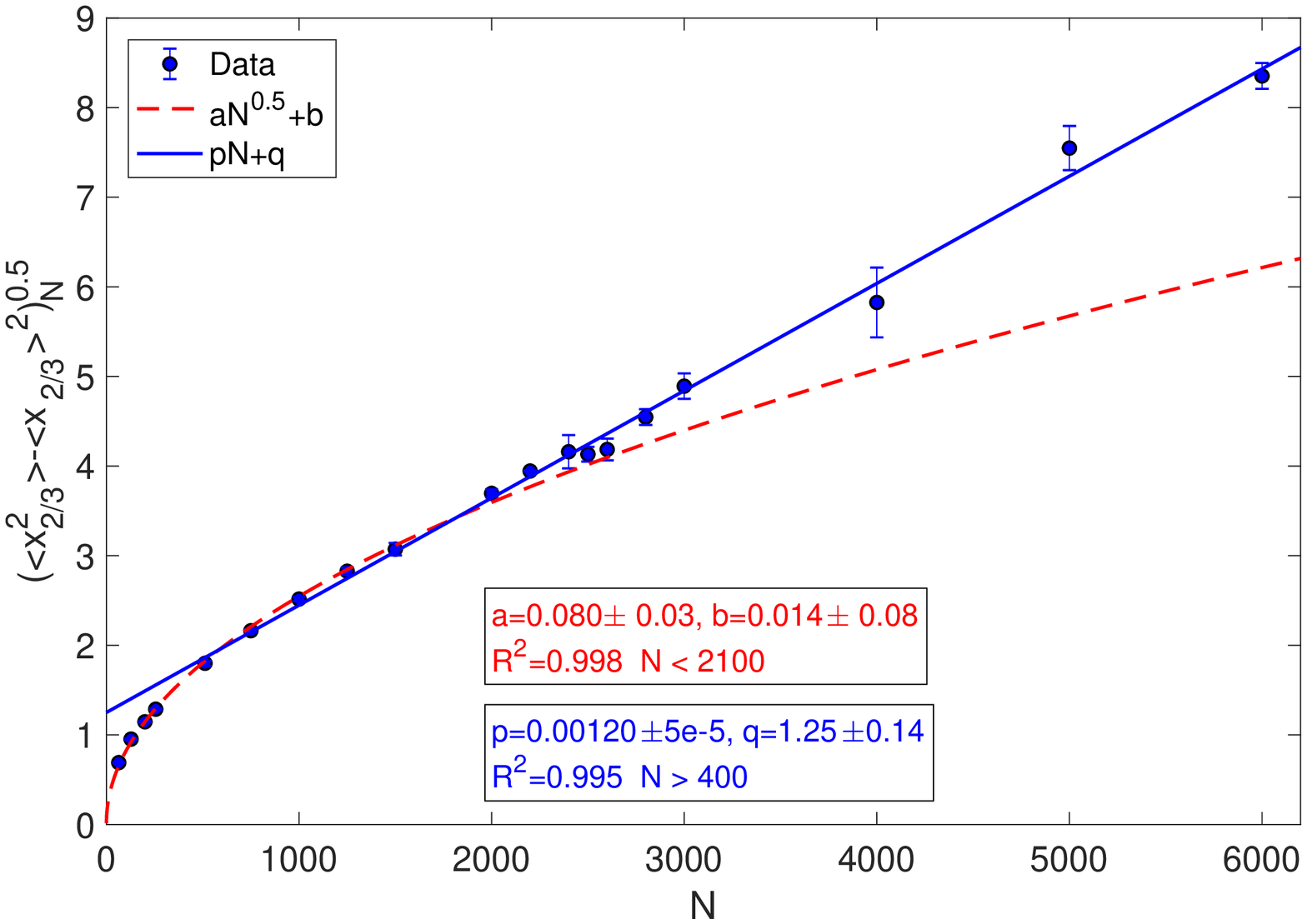}
\caption{\label{page89June} Upper panel: dependence on $N$ of the standard deviations of the
vibrations of particles at 1/3 of the chain.  Lower panel: dependence on $N$ of the standard 
deviations of the vibrations at 2/3 of the chain. In both cases, a square root and a linear 
fit are drawn. The square root fit holds at small $N$. At large $N$ the linear fit takes over.  
In both panels  $N$ ranges from 64 to 6000.
Particles motions look more like some kind of convection rather than like microscopic lattice
vibrations.}
\end{figure}

We find, however, that for every particle $i$, also the size of vibrations is ``macroscopic'':
$\sqrt{\langle x_i^2\rangle -\langle x_i \rangle^2} \sim O(N)$.
In the lower panel of Fig.\ref{page67June} and in Fig.\ref{page89June}, square root fits and linear 
fits are compared for $N$ ranging from 64 to 6000. 
The square root fits are appropriate for small $N$, while at large $N$ the linear 
fit takes over. The
size of these vibrations appears even more striking observing that displacing by a large amount 
one of them, a whole collection of particles must be correspondingly displaced. 
Indeed, the 
repulsive part of the LJ potential does not allow particles' order to be modified, as noted also in Ref.\cite{Politi}. 
Therefore, the motion of particles about their average positions is not an irregular motion about fixed 
positions. In accord with the observations on 
persistent correlations, this motion looks like a kind of convection, 
although LTE and standard hydrodynamics do not hold \cite{GibRon,NaRam02,MaNa06,Hurtado,Politi}.
It follows that, in these cases, energy transport cannot be directly related to ``heat'' flows. 
\begin{figure}
\hskip -20pt
\centering
\includegraphics[width=0.5\textwidth]{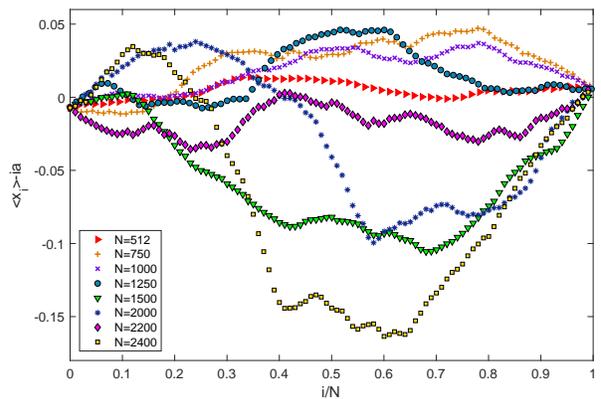} 
\caption{\label{MeanDispNoGrad} {Equilibrium simulations. Plot of the displacement
of the mean position of particle $i$ from its mechanical equilibrium position, $\langle x_i \rangle - i a$,  
for various values of $N$, for first and second neighbors interactions when $T_L=T_R=5$.
The deviations from the mechanical equilibrium are negligible.}}
\end{figure}
\begin{figure}
\includegraphics[width=0.5\textwidth]{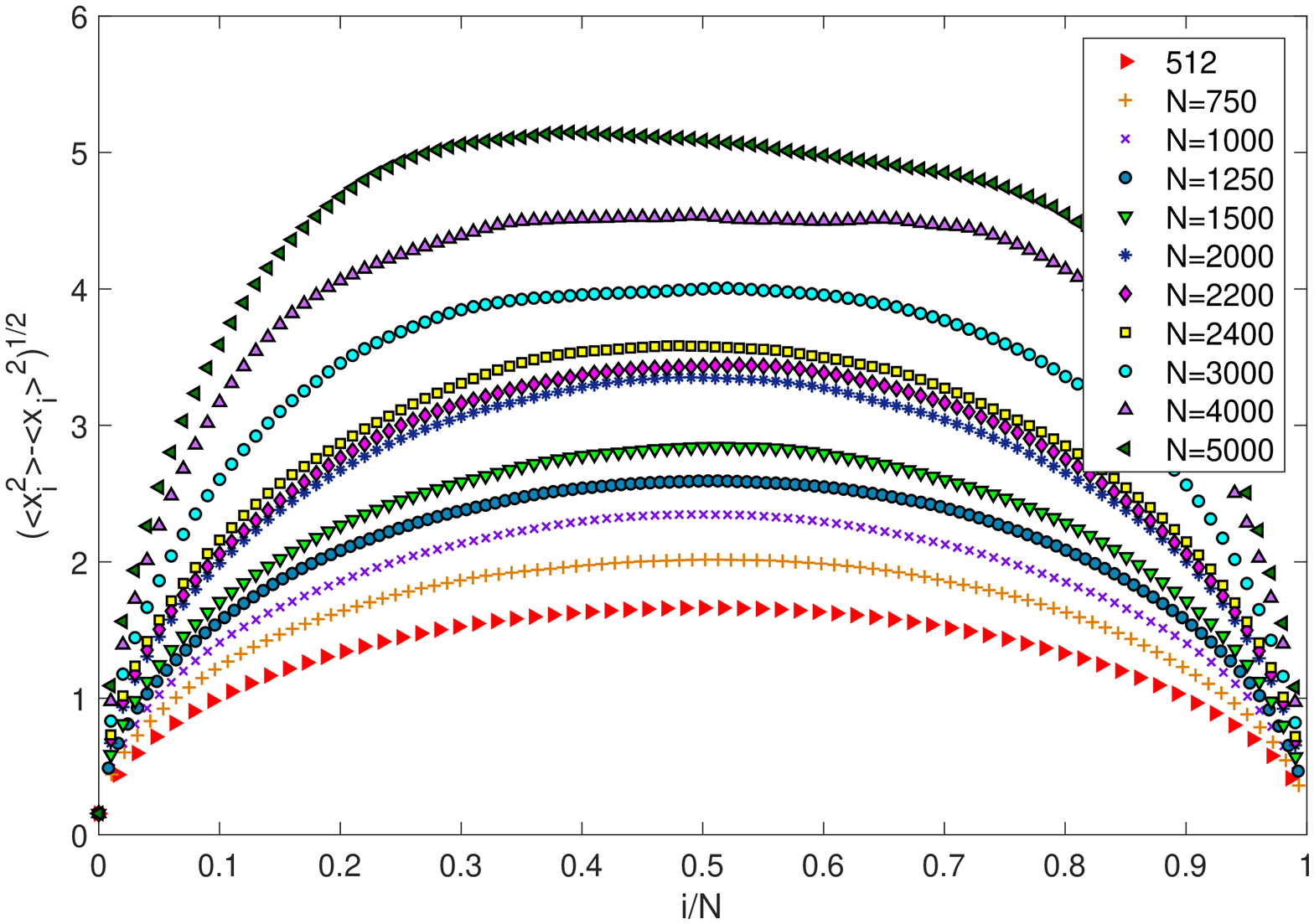} 
\includegraphics[width=0.5\textwidth]{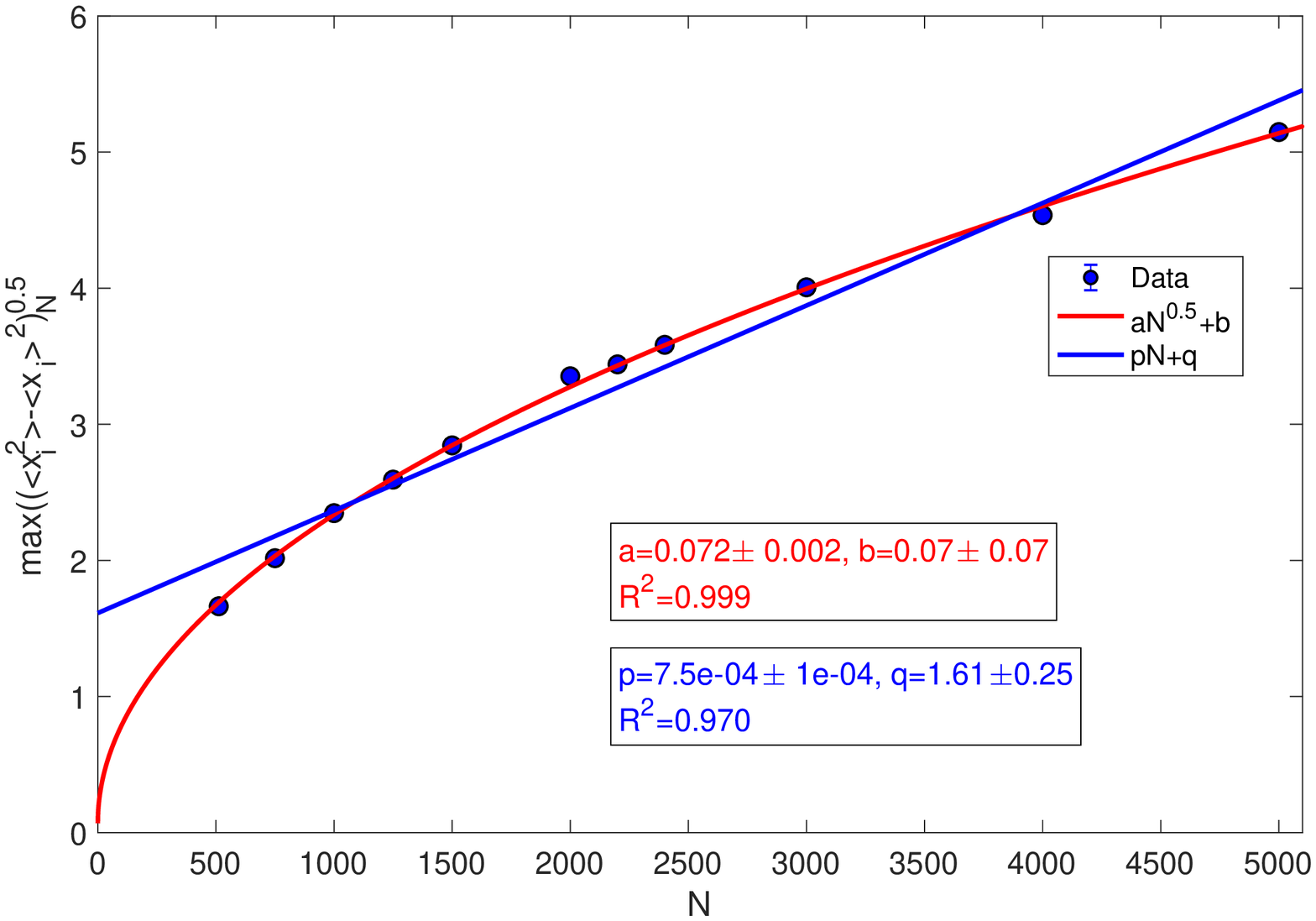}
\caption{\label{equilibrio} Equilibrium simulations $(T_L=T_R=5)$ for $N$ ranging from 512 to \cla{5000}. Upper panel: standard deviations of the particles vibrations 
about their average position $(x_i- \langle x_i \rangle)$, in lattice vectors units.
Lower panel: dependence on $N$ of the maximum standard deviation, together with linear  and  square root fits. 
This dependence on $N$ should not be confused with the $O(\sqrt{i})$ dependence on $i$ of Ref.\cite{Peierls}.}
\end{figure}

The situation is different for $T_L=T_R$. Figure \ref{MeanDispNoGrad} shows 
that the lattice deformations are much smaller than the lattice spacing $a$,
and can be neglected. The computed values of 
$(\langle x_i \rangle -ia)$ practically vanish and do not depend on $N$. 
The standard deviation of 
the vibrations about the mean position is represented in upper panel of 
Fig.\ref{equilibrio} and it appears to be closer to $O(\sqrt{N})$ than to $O(N)$ {as can be seen in lower panel of Fig.\ref{equilibrio}.
In this case, in which there is no net energy transport, the system also behaves more like a fluid than
like a solid in sense closer to that of \cite{Peierls}, although our results refers to a different situation.

\section{Heat flux}

\label{sec:heatflux}
In order to understand the effect of $O(N)$ fluctuations and lattice distortions on the
behaviour of usual microscopic quantities, let us consider the ``heat flux'' given
by Eq.(23) of Ref.\cite{LLP-PhyRep}. For the case of first and second nearest 
neighbors interactions, that expression must be modified as follows:
\be
\begin{split}
J_i=&\frac 1 2 (x_{i+1}-x_i)F_1(x_{i+1}-x_i)(\dot{x}_{i+1} + \dot{x}_i) \\
& +  (x_{i+2}-x_i)F_2(x_{i+2}-x_i)(\dot{x}_{i+2} + \dot{x}_i)+ \dot{x}_i h_i\, ,
\label{exactJ}
\end{split}
\ee
where $F_1$ and $F_2$ are defined by Eq.(\ref{F1F2}) and $h_i$ is the energy of the $i$-th particle.

The quantity $J_i$ is only apparently ``local'' because it quantifies a flow through the position of particle $i$, 
and not through a fixed position in space. Moreover, it implicitly
requires small position fluctuations and small lattice deformations, because Eq.(\ref{exactJ})
is obtained through Fourier analysis for spatially homogeneous systems, in the limit of small 
wave vectors, \cite{LLP-PhyRep,Dhar}. For instance, denoting by $k$ the wave-vector, Eq.(23) of 
Ref.\cite{LLP-PhyRep} follows from Eq.(21) only if $k(x_{n+1}-x_n)$ is small. 
On the contrary, in our cases, this quantity strongly varies in space and time, and average lattice
distortions are of order $O(N)$, cf.\ Section \ref{fluid}. Therefore, one 
expects $J_i$ to fail, and it is interesting to investigate how that 
is realized, varying the relevant model parameters.

For chains with nearest neighbors Lennard-Jones interactions ($F_2\equiv 0$ 
in Eq.(\ref{exactJ})),
we find that while the steady state heat flow should not depend on position, the time 
average of $J_i$ substantially changes with $i$, cf.\ Fig.\ref{flussi}. 
\begin{figure}
\includegraphics[width=0.5\textwidth]{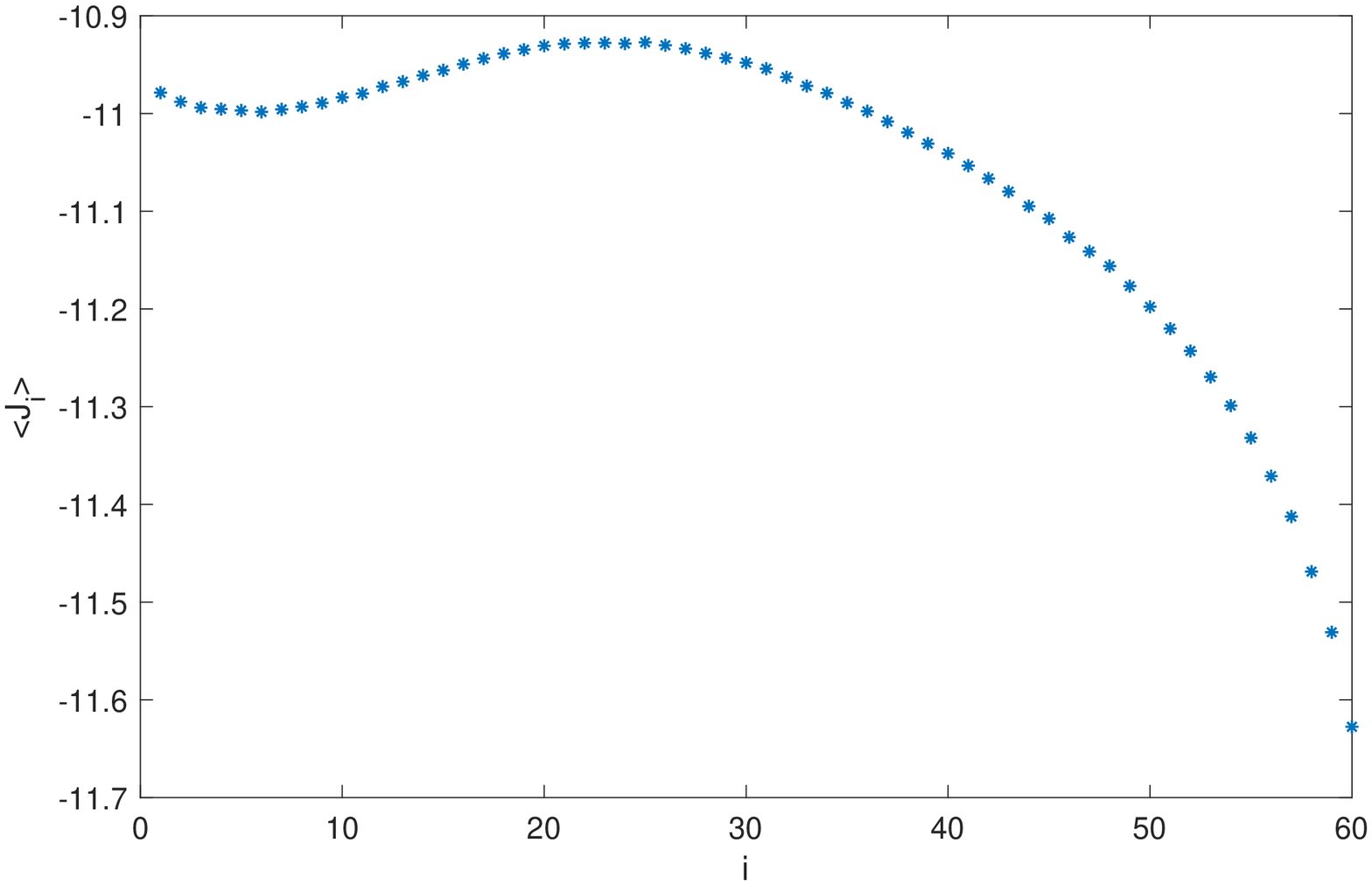} 
\includegraphics[width=0.5\textwidth]{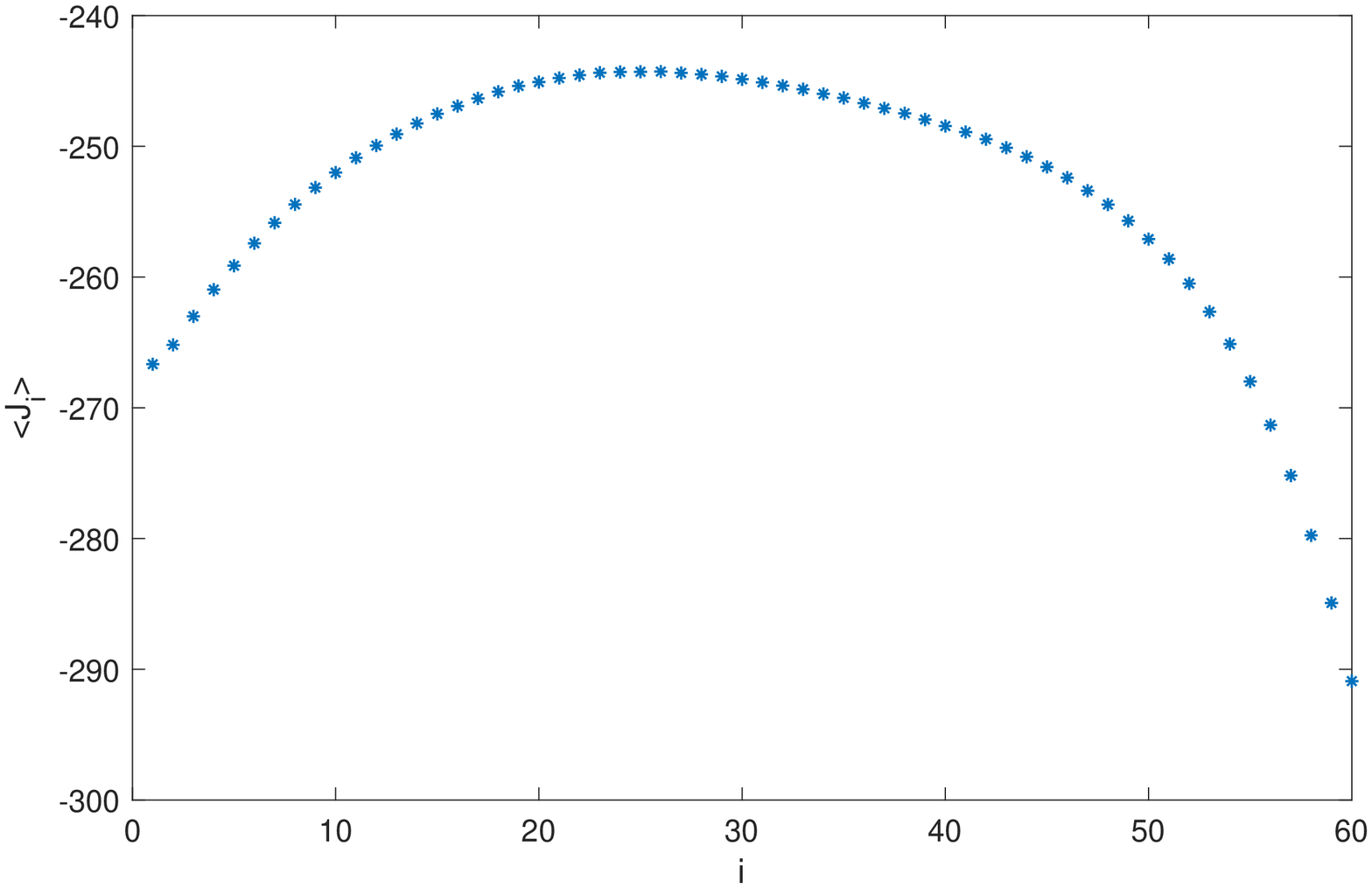} 
\caption{\label{flussi} 
Chains with nearest neighbors Lennard-Jones interactions.
Upper Panel: flux $\langle J_i \rangle$ computed according to (\ref{exactJ}), for $N=64$, $T_L=1$, $T_R=4$. 
Lower Panel: $\langle J_i \rangle$ for $N=64$, $T_L=1$, $T_R=64$.}
\end{figure}
To quantify this phenomenon, we introduce the relative variation of $\langle {J}_i \rangle$, 
$$
\delta= \left|{\max_i \langle {J}_{i}\rangle-\min_i \langle {J}_{i}\rangle \over \bar{J}} \right| ~, ~~\mbox{where } ~
\bar{J}={1 \over N} \sum_{i}\langle {J}_{i}\rangle ~ ,
$$ 
In Tables \ref{tab2bis} and \ref{tab4}, for average temperature gradients similar to those commonly found in the 
literature \cite{LLP,Kabu}, we observe that $\delta$ tends to grow with the temperature gradient, 
at fixed $N$. In general, however, reducing the average gradient by increasing the system size, does not lead to smaller 
$\delta$ \cite{observ}.

\begin{table}[htp]
\begin{center}
\begin{tabular}{|c|c|c|}
\hline
 $T_{R}$ & $\delta_1$ & $\delta_2$ \\
\hline
 1.1 & 0.0240  & 0.0199 \\
\hline
 1.5 & 0.0091 & 0.0077 \\
\hline
 2    &  0.0142 & 0.0145 \\
\hline
 4    &  0.0480 & 0.0481 \\
\hline
 8   &  0.0831 &  0.0829 \\
\hline
16  & 0.1060 &  0.1062 \\
\hline
32 & 0.1199  & 0.1201 \\
\hline
64 &  0.1229 &  0.1232\\
\hline
\end{tabular}
\end{center}
\caption{\label{tab2bis} Relative variation $\delta$ 
of the flux $J_i$ \cla{for}  $N=64$ particles with first and second nearest neighbors
interactions. $T_L=1$ while $T_R$ takes eight different values. $\delta_1$ is 
computed averaging over $2\cdot 10^9$ time steps, $\delta_2$ over $4 \cdot 10^9$ time steps. }
\end{table}

\begin{table}[htp]
\begin{center}
\begin{tabular}{|c|c|c|c|}
\hline
 $T_{R}$ & $N=64$ & $N=128$ & $N=256$ \\
\hline
 1.1 & 0.0240 & 0.0117 & 0.0110656  \\
\hline
 1.5 & 0.0091 & 0.0297 & 0.0317283  \\
\hline
 2    & 0.0142 & 0.0534 & 0.0555437 \\
\hline
 4    & 0.0480 & 0.0817  & 0.104345 \\
\hline
 8   &  0.0831 & 0.0659 &  0.0907829 \\
\hline
16  &  0.1060  & 0.0683 & 0.0485491  \\
\hline
32 &  0.1199 & 0.1560  &  0.0643797 \\
\hline
64 &  0.1229 & 0.2306 &  0.195046 \\
\hline
\end{tabular}
\end{center}
\caption{\label{tab4} Relative variation $\delta$ of the average fluxes $\langle J_i\rangle $ defined by Eq.(\ref{exactJ}).
Chains with $N=64$, $N=128$ and with $N=256$ particles, with nearest neighbors interactions are considered. 
Averages are computed over $2\cdot 10^9$ time steps. $T_L=1$, while $T_R$ takes eight different values.}
\end{table}
We conclude that under our conditions the quantity $J_i$ represents neither a heat nor an energy flow,
and that this is not a consequence of the size of temperature gradients, but of the size of fluctuations.
These increase with growing $N$, thus preventing LTE and standard hydrodynamics in the large $N$ limit \cite{Hurtado,GibRon,Politi}.
In the next section we propose a modification of $J_i$ that, taking into account 
the deformations of the lattice, is more stable than $J_i$ along the chain.

\section{Concluding remarks}
In this work we have presented numerical results concerning several kinds of 1D
systems of nonlinear oscillators, in contact with two Nos\'e-Hoover thermostats.
Scrutinizing the behaviour of mechanical quantities that are commonly considered 
in the specialized literature, we have investigated the fluctuations and lattice 
distortions, which are expected to prevent the establishment of ``thermodynamic'' 
regimes \cite{GibRon,JeppsR,JouBook,Politi}. 

\begin{figure}
\includegraphics[width=0.5\textwidth]{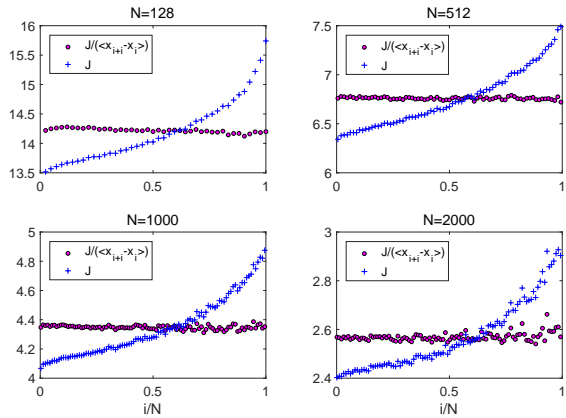}
\caption{\label{normJ}
Normalized energy flux $J^n$ (o) and the flux $J$ (+) defined by \eqref{exactJ} for chains of different lengths 
(N=128, 512, 1000, 2000) and with $T_L=1$ and $T_R=10$. 
Although $J^n$ 
is not exactly constant, at large $N$ it enjoys small fluctuations about a given average value.}
\end{figure}

Thermodynamic properties emerge indeed from
the collective behavior of very large assemblies of interacting particles, if correlations 
decay rapidly compared to observation time scales, and if boundary effects are negligible. 
While this is often the case of 3D mesoscopic cells containing large numbers of properly 
interacting particles, it is not obvious in 1D systems.

In particular, we have observed that temperature differences at the boundaries produce 
$O(N)$ fluctuations and deformations of the lattice, that result in strongly inhomogeneous systems. 
This should be taken into account when defining the heat conductivity. 
Furthermore, these $O(N)$ effects imply that larger $N$ is not going to make our systems closer 
to thermodynamic systems, when $N$ is increased, consequently standard hydrodynamics does not apply \cite{Hurtado,GibRon,Politi}.

In the light of the above observations, the definition of energy transport (\ref{exactJ}) may be 
modified in order to take into account the lattice deformations. One possibility that may considered
is to normalize the energy flow by the average distance between particles, introducing
\be
J^n_i = {J_i \over \langle x_{i+1} - x_i \rangle } ~, \qquad i = 2, ... , N-2\ .
\ee
The result, shown in Fig.\ref{normJ},
indicates that this avenue deserves further investigation, and will be the subject of future works.

\centerline{\bf ACKNOWLEDGMENTS } 
The authors are grateful to Carlos Mejia-Monasterio for extensive discussions
and enlightening remarks. The authors are grateful to Antonio Politi for very useful suggestions.
Antonio Politi, in particular has suggested the normalization of the energy flux.
This work is partially supported by Gruppo Nazionale per la
Fisica Matematica (GNFM-INdAM). C.G. and C.V. acknowledge financial supports
from  ``Fondo di Ateneo per la Ricerca 2016" and ``Fondo di Ateneo per la Ricerca 2017"-
Universit\ah\  di Modena e Reggio Emilia.


\begin{thebibliography}{1}
%

\bibitem{Lebow} Z.\ Rieder Z, J.L.\ Lebowitz, E.\ Lieb, 1967, {\em Properties of a harmonic crystal in a stationary nonequilibrium state},
J.\ Math.\ Phys.\ {\bf 8} 1073

\bibitem{ColGiaGibVer}  M.\ Colangeli, C.\ Giardin\ah , C.\ Giberti and C.\ Vernia, Phys. Rev. E {\bf 97}, 030103(R) (2018)

\bibitem{LandauV} L.D.\ Landau, E.M.\ Lifshitz, 1980, {\em Statistical Physics}
Volume 5 of Course of Theoretical Physics, Part 1, Pergamon Press, Oxford

\bibitem{Chibbaro} S.\ Chibbaro, L.\ Rondoni, A.\ Vulpiani, 2014, {\em Reductionism, Emergence and Levels of Reality}, Springer Verlag, New York

\bibitem{Prigogine} D.\ Kondepudi, I.\ Prigogine, 1998, {\em Modern thermodynamics : from heat engines to dissipative structures}, John Wiley \& Sons Ltd, Chichester

\bibitem{Spohn} H.\ Spohn, 1991 {\em Large Scale Dynamics of Interacting Particles}, Texts and
Monographs in Physics, Springer-Verlag, Heidelberg

\bibitem{Bellissard} J.\ Bellissard, {\em Coherent and Dissipative Transport in Aperiodic Solids: 
An Overview}, in P. Garbaczewski R. Olkiewicz (Eds.), 2002, {\em Dynamics of Dissipation}, Springer Verlag, Berlin

\bibitem{Kreuzer} H.J.\ Kreuzer, 1981, {\em Nonequilibrium thermodynamics and its statistical foundations}, Claredon Press, Oxford

\bibitem{Pigo1} M.\ Falcioni, L.\ Palatella, S.\ Pigolotti, L.\ Rondoni, A.\ Vulpiani, 2007, {\em Initial growth of Boltzmann entropy 
and chaos in a large assembly of weakly interacting systems}, Physica A {\bf 385} 170

\bibitem{Pigo2} L.\ Rondoni, S.\ Pigolotti, 2012, {\em On $\Gamma-$ and $\mu-$space descriptions: Gibbs and Boltzmann entropies of
symplectic coupled maps}, Phys. Scr. {\bf 86} 058513

\bibitem{DeGroot} S.R.\ de Groot and  P.\ Mazur 1984 {\em Non-equilibrium Thermodynamics}, Dover, New York

\bibitem{CR98}
E.G.D.~Cohen and L.~Rondoni, Chaos, {\bf 8}(2), 357 (1998)

\bibitem{RC02}
L.\ Rondoni and E.G.D.\ Cohen,
Physica D, {\bf 168-169}, 341 (2002)


\bibitem{AURIGA1} M.~Bonaldi \textit{et al.}, Phys.~Rev.~Lett. \textbf{103}, 010601 (2009)

\bibitem{AURIGA2} L.\ Conti, M.\ Bonaldi, L.\ Rondoni, 
CLASSICAL AND QUANTUM GRAVITY, {\bf 27} 084032  (2010)

\bibitem{Mish} 
Y. Mishin, Ann.\ Phys.\ {\bf 363} 48  2015

\bibitem{HickMish} 
J.\ Hickman and Y.\ Mishin, Phys.\ Rev.\ B {\bf 94} 184311 (2016)

\bibitem{StochTherm} U.\ Seifert, Rep.\ Prog.\ Phys.\ {\bf 75} 126001 (2012)

\bibitem{CasatiRev}
 G.\ Benenti, G.\ Casati, K.\ Saito and  R.S.\ Whitney,
arXiv:1608.05595 [cond-mat.mes-hall] (2016)

\bibitem{Li1} Zhibin Gao, Nianbei Li, Baowen Li, Phys. Rev. E {\bf 93}, 022102 (2016). 

\bibitem{Dhar}
S.G.\ Das,  A.\ Dhar, O.\ Narayan, 
J.\ Stat.\ Phys. {\bf 154} 204  (2014) 

\bibitem{Li2}
S.\ Liu, P.\ H\"anggi, N.\ Li, Jie Ren, B.\ Li,
Phys. Rev. Lett. 112, 040601 (2014)

\bibitem{LepriEd} S.Lepri (Ed.) 2016 {\em Thermal  transport in low dimensions - From Statistical Physics to Nanoscale Heat Transfer}, Lecture Notes in Physics, vol. {\bf 921}, Springer, Heidelberg

\bibitem{CasatiJ} J.\ Wang, G.\ Casati, arXiv:1610.07474 [cond-mat.stat-mech] (2016)

\bibitem{ExpAnomFourier} C. W. Chang, D. Okawa, H. Garcia, A. Majumdar, A. Zettl, Phys.\ Rev. Lett.\ {\bf 101} 075903 \ (2008);
X.\ Xu et al., Nature Comm.\ 5:3689, 1 (2014)

\bibitem{JeppsR} O.G.\ Jepps and L. Rondoni, 
J.\ Phys.\ A {\bf 39}, 1311 (2006) 

\bibitem{GibRon} C.\ Giberti and L.\  Rondoni, Phys.\ Rev.\ E {\bf 83}, 041115 (2011)

\bibitem{Zhao1}
S.\ Chen, Y.\ Zhang, J.\ Wang, H.\ Zhao, J. Stat. Mech.  033205 (2016)

\bibitem{Zhao2} S.\ Chen, Y.\  Zhang, J.\ Wang, H.\ Zhao, Phys. Rev. E {\bf 87}, 032153 (2013)

\bibitem{onorato} M.\ Onorato, L.\ Vozella, D.\ Proment, Y.V.\ Lvov, www.pnas.org/cgi/doi/10.1073/pnas.1404397112

\bibitem{Saito} A.\ Dhar and K.\ Saito,
Phys.\ Rev.\ E {\bf 78} 061136 (2008)

\bibitem{Peierls} R.E.\ Peierls, {\it Surprises in theoretical physics}, Princeton University Press, Princeton, New Jersey (1979), Sec.4.1.

\bibitem{NaRam02} O.\ Narayan and S.\ Ramaswamy, {\it Anomalous heat conduction
in one-dimensional momentum-conserving systems}, Phys.\ Rev.\ Lett.\ {bf 89} 200601 (2002)

\bibitem{MaNa06} T.\ Mai and O.\ Narayan, {\it Universality of one-dimensional heat
conductivity}, Phys.\ Rev.\ E {\bf 73} 061202 (2006)

\bibitem{Breaking} L.\ Conti et al, 2013, {\em Effects of breaking vibrational energy
equipartition on measurements of temperature in macroscopic oscillators subject to heat flux}, J.\ Stat.\ Mech.\ P12003

\bibitem{Sengers} J.M.\ Ortiz de Zárate, J.V.\ Sengers, 2006, {\em  Hydrodynamic Fluctuations in Fluids and Fluid Mixtures}, 
Elsevier, Amsterdam

\bibitem{PSV17} A.\ Puglisi, A.\ Sarracino, A.\ Vulpiani 2017 {\em Temperature in and out of equilibrium: A review of 
concepts, tools and attempts} Phys.\ Rep.\ {\bf 709-710}, 1

\bibitem{EWSR2016} D.J.\ Evans, S.R.\ Williams, D.J.\ Searles, L.\ Rondoni, 2016, {\em On Typicality in Nonequilibrium Steady States},
J.\ Stat.\ Phys.\ {\bf 164} 842

\bibitem{GGnoneq} G.\ Gallavotti, 2014, {\em Nonequilibrium and Irreversibility}, Springer, New York

\bibitem{MR} G.P.\ Morriss, L.\ Rondoni, Phys.\ Rev.\ E {\bf 59} R5 (1999)

\bibitem{Jou1} J.\ Casas-Vázquez, D.\ Jou, Rep.\ Prog.\ Phys.\ {\bf 66} 1937 (2003)

\bibitem{Jou2} D.\ Jou, L.\ Restuccia, Physica A {\bf 460} 246 (2016)

\bibitem{He} X.\ Cao and D.\ He, Phys.\ Rev.\ E {\bf 92} 032135 (2015)

\bibitem{JouBook} D.J.\ Jou, J.\ Casas-V\`azquez, G.\ Lebon, 2010, {\em Extended Irreversible Thermodynamics}, Springer, New York

\bibitem{Criado} M.\ Criado-Sancho, D.\ Jou, J.\ Casas-Vázquez, 2006, {\em Nonequilibrium kinetic temperatures in flowing
gases}, Phys.\ Lett.\ A {\bf 350} 339

\bibitem{Zema} M.W.\ Zemansky and R.H.\ Dittman 1997 {\em Heat and Thermodynamics}, McGraw-Hill, New York

\bibitem{inpreparat} C.\ Giberti, L.\ Rondoni, C.\ Vernia (in preparation)

\bibitem{LLP-PhyRep} S.\ Lepri, R.\ Livi and A.\ Politi, Phys.\ Rep.\ {\bf 377}, 1-80 (2003)


\bibitem{Hurtado} P.I.\ Hurtado, {\it Breakdown of hydrodynamics in a simple one-dimensional
fluid}, Phys.\ Rev.\ Lett.\ {\bf 96} 010601 (2006)

\bibitem{Politi} S.\ Lepri, P.\ Sandri, A.\ Politi, Eur. Phys. J. B {\bf 47}, 549-555 (2005)

\bibitem{1st2ndpaper} P.\ De Gregorio, L.\ Rondoni, M.\ Bonaldi and L.\ Conti, Phys.\ Rev.\ B
{\bf 84}, 224103 (2011). L.\ Conti, P.\ De Gregorio, M.\  Bonaldi, A.\ Borrielli, M.\ Crivellari, 
G.\ Karapetyan, C.\ Poli, E.\ Serra, R.K.\ Thakur and L.\ Rondoni
Phys.\ Rev.\ E {\bf 85}, 066605 (2012)

\bibitem{InSome} In some  
cases, we extended the Lennard-Jones interaction to the third nearest neighbors, preserving the equilibrium 
configuration $x_i=i a$. The corresponding equations of motion and thermostats are the natural modification of 
the previous ones, hence are not reported here.

\bibitem{Virial} In Ref.\cite{Falasco} a nonequilibrium mesoscopic version of the virial relation in given.


\bibitem{MejiaPoliti} L.\ Delfini, S.\ Lepri, R.\ Livi, A.\ Politi, Phys.\ Rev.\ E {\bf 73},
060201(R) (2006). S.\ Lepri, C.\ Mej\'{i}a-Monasterio, A.\ Politi, J.\ Phys.\ A
{\bf 43}, 065002 (2010); L.\ Delfini, S.\ Lepri, R.\ Livi, C.\ Mej\'{i}a-Monasterio, A.\ Politi, 
ibid. {\bf 43}, 145001 (2010).

\bibitem{LLP}  S.\ Lepri, R.\ Livi and A.\ Politi, Phys. Rev. Lett., {\bf 78} (10), 1896--1899 (1997); S.\ Lepri, R.\ Livi and A.\ Politi, Phsica D {\bf 119}, 140--147 (1998)

\bibitem{DharReview} A.\ Dhar,   {\em Heat Transport in low-dimensional systems } Advances in Physics, Vol. 57, No. 5, 457-537 (2008)

\bibitem{RJepps} S.J.\ Davie, O. G.\  Jepps, L.\ Rondoni, J. C.\ Reid, D. J.\  Searles,  Phys.\ Script.\ {\bf 89}, 048002 (2014)

\bibitem{BiancaJR} O.\  Jepps, C.\ Bianca and  L.\ Rondoni  CHAOS {\bf 18}, 013127 (2008).  C.\ Bianca and L.\ Rondoni  CHAOS {\bf 19}, 013121 (2009)

\bibitem{Salari} L.\ Salari, L.\ Rondoni, C.\ Giberti, R.\ Klages, CHAOS
{\bf 25}, 073113 (2015)

\bibitem{Kabu} H.\ Kaburaki, M.\ Machida, Phys.\ Lett.\ A {\bf 181}  85 (1993)

\bibitem{observ} Actually, for mere energy flows, there is no reason to be bounded by small 
temperature gradients.

\bibitem{Falasco} G.\ Falasco, F.\ Baldovin, K.\ Kroy, M.\ Baiesi, New J.\ Phys.\ {\bf 18} 093043 (2016)







%
%



















\end{thebibliography}
\end{document}